# Dynamical symmetry of strongly light-driven electronic system in crystalline solids

Kohei Nagai[1,*], Kento Uchida[1], Naotaka Yoshikawa[1], Takahiko Endo[2], Yasumitsu Miyata[2], and Koichiro Tanaka[1,†]

[1]Department of Physics, Graduate School of Science, Kyoto University, Sakyo-ku, Kyoto 606-8502, Japan

[2]Department of Physics, Tokyo Metropolitan University, Hachioji, Tokyo 192-0397, Japan

* nagai.kohei.38w@st.kyoto-u.ac.jp

† kochan@scphys.kyoto-u.ac.jp

**The Floquet state, which is a periodically and intensely light driven quantum state in solids, has been attracting attention as a novel state that is coherently controllable on an ultrafast time scale**[1–7]**. An important issue has been to demonstrate experimentally novel electronic properties in the Floquet state. One technique to demonstrate them is the light scattering spectroscopy, which offers an important clue to clarifying the symmetries and energy structures of the states through symmetry analysis of the polarization selection rules. Here, we determine circular and linear polarization selection rules of light scattering in a mid-infrared-driven Floquet system in monolayer $MoS_2$ and provide a comprehensive understanding in terms of the "dynamical symmetry" of the Floquet state.**

Floquet engineering is a potential concept for coherent control of electronic states under a strong light field[1,2]. The Floquet theoretical approach is useful for describing strong light-matter interactions at energy scales beyond which perturbation theory works[8]. In this nonperturbative regime, intense light is predicted to change the symmetry and topology of the states and in turn the electronic properties of solids[1,2].

The Floquet state in solids has been verified through time and angle-resolved photoemission spectroscopy (Tr-ARPES)[3,5], time-resolved absorption spectroscopy[4,6], and time-resolved transport measurements[7]. Its properties, such as nonperturbative electron dynamics during the period of the driving laser, can also be explored by using high-order harmonic generation (HHG)[13–15], which is a coherent emission process from a Floquet system[9–12].

Here, we examine light scattering in a Floquet system by injecting an additional probe pulse. Compared with HHG, tuning of the polarization and frequency of the probe light may provide more detailed information about the symmetries and electronic structures. This process is nothing but high order sideband generation (HSG)[16–19].

Below, we systematically present polarization selection rules, which is fundamental to probe the symmetry of the electronic states, for HSG in monolayer $MoS_2$ under a mid infrared (MIR) driving field. In a Floquet system, the electronic properties are described by a unique class of symmetries, called "dynamical symmetries" (DSs), which unify the symmetries of the spatio-temporal profiles of the laser field and material[3,9,11,12,20,21]. It has been experimentally confirmed that DSs govern the band crossings of surface electrons in a light driven topological insulator[3] and determines the polarization selection rules for HHG in a circularly polarized light driven crystalline solid[12] Here, we introduce a new interpretation, i.e., HSG as "Raman scattering" of the MIR-driven Floquet state (Fig. 1(a)), and use the DS concept to achieve a full understanding of the polarization selection rules.

We prepared monolayer $MoS_2$ grown with the CVD method (see Sample Preparation). This atomically thin semiconductor is an ideal experimental HSG platform on which to avoid propagation effects[14,15,22–25]. Figure 1 (b) shows a schematic diagram of the HSG measurement setup. We used intense MIR pulses (photon energy: $\hbar\omega_{MIR}$ =0.26 eV, pulse duration: 60 fs) to create a Floquet state in monolayer $MoS_2$. To achieve a nonperturbative regime without damaging the monolayer, we set the photon energy of the pulses to a much lower energy than the bandgap energy of the monolayer [15] (1.8 eV). In addition, we injected weak near infrared (NIR) pulses nearly resonant with the bandgap energy (photon energy: $\hbar\omega_{NIR}$=1.55 eV, pulse duration: 110 fs) into the MIR-driven system. We controlled the polarizations of the MIR and NIR pulses by using liquid crystal retarders and resolved the polarization of the sidebands by using wave plates and polarizers. The sideband spectra were detected by a spectrometer equipped with a CCD camera (see Methods). Throughout this paper, we denote the zigzag direction of monolayer $MoS_2$ as X and armchair direction as Y.

We observed HHG spectra at photon energies of $m\hbar\omega_{MIR}$ ($m$ : integer) by irradiating the monolayer with MIR pulses and observed the HSG spectra at photon energies of $\hbar\omega_{NIR} + m\hbar\omega_{MIR}$ ($m$: integer) by simultaneously applying NIR pulses. Therefore, we obtained the HSG contribution by subtracting the HHG component from the spectra (Supplementary Information 1). Figure 2 shows polarization-unresolved HSG spectra under linearly and circularly polarized excitation with a MIR-pulse peak intensity of 0.5 TW/cm$^2$ and NIR-pulse peak intensity of 0.5 GW/cm$^2$. Here, sidebands up to seventh order appear under linearly polarized excitations. In contrast, sidebands only up to third order appear under circularly polarized excitation. This difference may arise from the different resultant kinetic energies that coherent electron-hole pairs obtain from linearly and circularly polarized laser fields[17,19].

Nonperturbative aspects induced by MIR light appear in both the spectral shape and excitation power dependence. One such aspect is the non-exponential decay with increasing order in the spectra. The excitation power dependence is shown in Fig. S2. Moreover, the MIR power dependence deviates from the power-law derived from perturbation theory under both linearly and circularly polarized excitations (Fig. S2 **a,b**). On the other hand, the intensity of the sideband is proportional to the NIR probe power (Fig. S2 **c,d**). This indicates the NIR pulses can be treated as a perturbation.

The observed selection rules are shown in Fig. 3. Figures **a-d** shows circular ($\sigma^+$, $\sigma^-$) polarization-resolved sideband spectra obtained from different combinations of $\sigma^+$- and $\sigma^-$-polarized excitations. The polarization of the sideband depends on the order and the polarization of the excitation pulses. Since we obtained the same result from monolayer $MoSe_2$, which has the same crystal structure, the selection rules are determined only by the symmetry of the crystal and polarization of the light (Fig. S3).

Furthermore, we systematically examined the linear polarization selection rules. Figure 3 e-h shows linear polarization-resolved sideband spectra obtained from different combinations of X and Y-polarized excitations. In particular, when MIR driving pulses are X-polarized (Fig. 3 e, f), odd-order sidebands are emitted with a perpendicular polarization to that of NIR pulses and even-order sidebands are emitted with a parallel polarization. On the other hand, when the MIR driving pulses are Y-polarized (Fig. 3 g, h), the polarization of the sideband is parallel to that of NIR pulses for all orders.

To explain these selection rules, we propose a simple scheme for symmetry analysis of HSG using the "Raman tensor" and DSs. In the previous report, HSG selection rules were explained in terms of the symmetry of microscopic intraband dynamics of electron hole pairs in momentum space[19]. However, it is difficult to extend this microscopic explanation to the circular polarization case or to polarization selection rules in other materials.

Below, we show that DS gives us a general tool for symmetry analysis of HSG. To justify the "Raman-scattering" description, we consider a microscopic model of HSG. We start with a general Hamiltonian of an electron interacting with an electric field in a solid. We apply the Floquet theorem by assuming periodicity of the MIR driving field and perturbation theory for the weak NIR pulses (Supplementary Information 2)[8]. We calculated the polarization current in the MIR-driven Floquet system perturbed by additional NIR pulses, which emits sideband light as follows:

$$J_{SG,\mu}(t) = -\frac{i}{\hbar} \sum_{\nu} \sum_{e_F \neq i_F} \int_{-\infty}^{t} dt' e^{-i\omega_{ie}(t-t')} \langle i_F(r,t)|\hat{J}_\mu|e_F(r,t)\rangle$$

$$\times \langle e_F(r,t')|\hat{J}_\nu|i_F(r,t')\rangle \cdot \delta A_{NIR,\nu}(t') + c.c.. \quad (1)$$

Here, $\hat{J}_\mu = \partial \hat{H}_0(r,t)/\partial A_\mu(t)$ is the current operator in the velocity gauge, $\hat{H}_0(r,t)$ is the Hamiltonian of the system of a crystalline solid and external MIR laser field, $|i_F\rangle$ and $|e_F\rangle$ are initial and intermediate Floquet eigenstates, $\hbar\omega_{ie} = \hbar\omega_i - \hbar\omega_e$ is the difference in quasienergy between each Floquet eigenstate and $\delta A_{NIR,j}$ is the vector potential of the additional NIR pulses (see Supplementary Information 2). Equation (1) justifies the "Raman scattering" process description depicted in Fig. 1(a). A NIR photon coherently excites an electron from the initial to an intermediate Floquet state, and the electron simultaneously relaxes back to the initial state by emitting a sideband photon. In this interpretation, the high-order sideband corresponds to "multi-photon anti-Stokes Raman scattering". The second rank tensor $\langle i_F(r,t)|\hat{J}_\mu|e_F(r,t)\rangle \times \langle e_F(r,t')|\hat{J}_\nu|i_F(r,t')\rangle$ corresponds to the response function of the Floquet system, which gives the relation between $\boldsymbol{J}_{SG}(t)$ and $\delta \boldsymbol{A}_{NIR}(t)$.

This relation is restricted by the DS of the MIR-driven Floquet system. One way to describe this restriction is to consider a second rank tensor $\boldsymbol{J}_{SG}(t)\delta \boldsymbol{A}_{NIR}^\dagger(t)$. One can derive the invariance of this tensor under the DS operation when $\delta \boldsymbol{A}_{NIR}(t)$ is an "eigenvector" of the DS operation (Supplementary Information 3). $\boldsymbol{J}_{SG}(t)$ and $\delta \boldsymbol{A}_{NIR}(t)$ are directly related to the electric fields of the $m$-th order sideband $\boldsymbol{E}_{SG,m}(t)$ and the NIR light $\boldsymbol{E}_{NIR}(t)$. Therefore, the symmetry restriction can be written in terms of a "Raman tensor" as follows:

$$\mathcal{R}_m(t) = \boldsymbol{E}_{SG,m}(t)\boldsymbol{E}_{NIR}^\dagger(t)$$

$$= \begin{pmatrix} E_{SG,m,x}(t)E^*_{NIR,x}(t) & E_{SG,m,x}(t)E^*_{NIR,y}(t) \\ E_{SG,m,y}(t)E^*_{NIR,x}(t) & E_{SG,m,y}(t)E^*_{NIR,y}(t) \end{pmatrix}, \quad (2)$$

where $E_{SG,m,x}$ and $E_{NIR,x}$ ($E_{SG,m,y}$ and $E_{NIR,y}$) denote the X (Y) component of the electric fields of the $m$-th order sideband and NIR light, respectively. Here, the direction of the electric field is reduced to being in a two-dimensional space parallel to the monolayer sample in our experimental setup. The "Raman tensor" satisfies the DS of the Floquet system:

$$\hat{X}\mathcal{R}_m(t) = \mathcal{R}_m(t), \quad (3)$$

where $\hat{X}$ is a DS operation. Equation (3) determines the polarization selection rules of HSG. This is similar to the conventional Raman scattering in crystalline solids, where the tensor has the symmetry of the solid and determines the selection rules.

The DS operation of the Floquet system is determined by the symmetry of the monolayer MoS$_2$ and the MIR driving field. Thus, the DS of the system under circularly polarized MIR light is different from that under linearly polarized MIR light. Under circularly polarized light, the DS operator that determines the selection rule is

$$\hat{C}_{3,\sigma_M} = \hat{\tau}_{-3\sigma_M} \cdot \hat{R}_3, \quad (4)$$

where $\sigma_M = \pm 1$ is the polarization of the MIR driving pulses, $\hat{\tau}_n$ is a temporal translation by $T/n$ ($T$ : the period of the MIR light field) and $\hat{R}_3$ is a spatial rotation by $2\pi/3$. The angle of $2\pi/3$ reflects the three-fold rotational symmetry of the crystal (Supplementary Information 4). From equation (3) and (4), the circular polarization selection rule of the $m$-th order sideband is derived as

$$m\sigma_{MIR} + \sigma_{NIR} - \sigma_S^m = 3N, \quad (5)$$

where $\sigma_{NIR}, \sigma_S^m = \pm 1$ denotes the polarization of the NIR and $m$-th order sideband, respectively, and $N$ is an integer (Supplementary Information 5). Equation (5) is consistent with all experimental results in Fig. 3 **a-d**.

Equation (5) represents the angular momentum conservation rule of light modified in crystalline solids[20,26,27]. The $\sigma^+$ ($\sigma^-$)-polarized light can be considered to have spin $+\hbar$ ($-\hbar$). The left-hand side of equation (5) shows the difference between the total spin of the incident photons and the spin of the emitted $m$-th order sideband photon. Although the right-hand side should be zero in isotropic media, $3N\hbar$ is allowed in monolayer MoS$_2$, which has three-fold rotational symmetry, due to the rotational analogue of the umklapp process.

The linear polarization selection rule can be derived similarly. The DS of the Floquet system depends on whether the polarization of the MIR pulses is in the X or Y direction, reflecting the mirror symmetry with respect to the Y direction of monolayer MoS$_2$. The following DS operations determine the selection rules, respectively:

$$\hat{Z}_y = \hat{\tau}_2 \cdot \hat{\sigma}_y \tag{6}$$

$$\hat{\sigma}_y, \tag{7}$$

where $\hat{\sigma}_y$ means reflection with respect to the Y direction. Because of the temporal term in equation (6), the polarizations of the odd- and even-order sidebands are perpendicular to each other in the X-polarized case. On the other hand, all sidebands have the same polarization in the Y-polarized case (Supplementary Information 4,5). This is consistent with the experimental results in Fig. 3 **e-h**. Table 1 summarizes the polarization selection rules in each configuration.

Odd-order sidebands in monolayer WSe$_2$ with Y-polarized MIR pulses were not observed in the previous report, although they were observed in our experiment[19]. This fact does not contradict our selection rules. This difference may be attributed to the resonant condition between the NIR light and the ladder-like energy levels in the MIR-driven Floquet state.

In conclusion, we have systematically determined the circular and linear polarization selection rules of HSG in monolayer MoS$_2$. By combining the concepts of Floquet and perturbation theory, we devised a new description of HSG as a "Raman scattering" in the MIR-driven Floquet state and revealed that the selection rules of HSG can be comprehensively understood in terms of DS. DS has the potential to describe topological phases and classify Floquet systems such as Floquet topological insulators[2,21]. Thus, our results pave the way for experimental studies of electronic structures and their topological properties of Floquet systems through light-scattering experiments and DS analyses.

# Methods

## Sample preparation

Monolayer MoS$_2$ and MoSe$_2$ were grown on sapphire substrates by chemical vapor deposition. The monolayer flake size of MoS$_2$ was typically one hundred micrometers and MoSe$_2$ was typically tens of micrometers. The MoS$_2$ monolayers were prepared by using the method reported in ref. 28. The monolayer MoSe$_2$ was purchased from 2d Semiconductors, Inc.

## Experiments

The experimental setup for the HSG measurements is shown in Fig. S1. Ultrafast laser pulses (photon energy 1.55 eV, 35 fs pulse duration, 1 kHz repetition rate, 7 mJ pulse energy) were derived from a Ti:sapphire based regenerative amplifier. Some of the total pulse energy (1 mJ) was used to generate signal and idler beams by an optical parametric amplifier (OPA, TOPAS-C, Light Conversion). The MIR pulses (photon energy 0.26 eV) were obtained by difference frequency generation (DFG) of the signal and idler beams in a AgGaS$_2$ crystal. After the DFG, the signal and idler beams were blocked by a longpass filter (LPF) with a cutoff wavelength of 4 μm. Another part of the ultrafast laser pulses was passed through a band pass filter (BPF) centered at 800 nm (band width 10 nm) and used as the NIR probe pulses. The polarizations of the NIR and MIR pulses were controlled by a wire grid polarizer (WGP), a Glan laser polarizer (GLP), and liquid crystal variable retarders (LCR). The MIR pulses were focused by a ZnSe lens to a spot 60 μm in diameter (full width at half maximum). The NIR pulses were passed through a fused-silica lens and reflected by a D-shaped mirror placed below the MIR beam. The NIR pulses were focused onto the sample almost coaxially with the MIR beams (approximately 4 degrees between the two beams). The pulse durations of NIR and MIR pulses were estimated to be 110 fs and 60 fs (full width half maximum (FWHM)), respectively, as noted in Supplementary Information 6. The generated harmonics and sidebands were collected by fused-silica lens and their spectra were analyzed by a grating spectrometer (iHR320, Horiba) equipped with a Peltier-cooled Si charge-coupled device camera (Syncerity CCD, Horiba). The NIR light was blocked by 750 nm and 550 nm short pass filter (SPF) in front of the spectrometer. The polarization of the sidebands was resolved by a quarter wave plate and a wire grid polarizer. The obtained spectra were corrected for the total efficiency including mirrors, a spectrometer and a CCD camera. All the experiments were performed in the air at room temperature.

**Acknowledgement.**

This work was supported by a Grant-in-Aid for Scientific Research (S) (Grant No. 17H06124). K.U. was supported by a Grant-in-Aid for Early-Career Scientists (Grant No. 19K14632). N.Y. was supported by a JSPS fellowship (Grant No. 16J10537). Y.M. acknowledges support from JST CREST (Grant No. JPMJCR16F3) and a Grant-in-Aid for Scientific Research (B) (Grant No. JP18H01832) from the Ministry of Education, Culture, Sports, Science and Technology (MEXT), Japan.

**Author Contributions**

K.N., N.Y., and K.T. conceived the experiments. K.N., K.U., and N.Y. built the experimental setup. K.N. and N.Y. carried out the experiments. T.E. and Y.M. fabricated the samples. K.N., K.U., and K.T. constructed the theoretical framework and wrote the manuscript. All the authors contributed to the discussion and interpretation of the results.

**Competing interests**

The authors declare no competing interests.

**References**

1. Oka, T. & Aoki, H. Photovoltaic Hall effect in graphene. *Phys. Rev. B* **79**, 1–4 (2009).
2. Lindner, N. H., Refael, G. & Galitski, V. Floquet topological insulator in semiconductor quantum wells. *Nat. Phys.* **7**, 490–495 (2011).

3.  Wang, Y. H., Steinberg, H., Jarillo-Herrero, P. & Gedik, N. Observation of Floquet-Bloch states on the surface of a topological insulator. *Science* **342**, 453–457 (2013).

4.  Sie, E. J. *et al.* Valley-selective optical Stark effect in monolayer $WS_2$. *Nat. Mater.* **14**, 290–294 (2015).

5.  Mahmood, F. *et al.* Selective scattering between Floquet-Bloch and Volkov states in a topological insulator. *Nat. Phys.* **12**, 306–310 (2016).

6.  Uchida, K. *et al.* Subcycle Optical Response Caused by a Terahertz Dressed State with Phase-Locked Wave Functions. *Phys. Rev. Lett.* **117**, 277402 (2016).

7.  McIver, J. W. *et al.* Light-induced anomalous Hall effect in graphene. *Nat. Phys.* **16**, 38–42 (2019).

8.  Shirley, J. H. Solution of the Schrödinger equation with a Hamiltonian periodic in time. *Phys. Rev.* **138**, B979 (1965).

9.  Ben-Tal, N., Moiseyev, N. & Beswick, A. The effect of Hamiltonian symmetry on generation of odd and even harmonics. *J. Phys. B At. Mol. Opt. Phys.* **26**, 3017–3024 (1993).

10. Faisal, F. H. M. & Kamiński, J. Z. Floquet-Bloch theory of high-harmonic generation in periodic structures. *Phys. Rev. A* **56**, 748–762 (1997).

11. Alon, O., Averbukh, V. & Moiseyev, N. Selection rules for the high harmonic generation spectra. *Phys. Rev. Lett.* **80**, 3743–3746 (1998).

12. Neufeld, O., Podolsky, D. & Cohen, O. Floquet group theory and its application to selection rules in harmonic generation. *Nat. Commun.* **10**, 405 (2019).

13. Ghimire, S. *et al.* Observation of high-order harmonic generation in a bulk crystal. *Nat. Phys.* **7**, 138–141 (2011).

14. Yoshikawa, N., Tamaya, T. & Tanaka, K. High-harmonic generation in graphene enhanced by elliptically polarized light excitation. *Science* **356**, 736–738 (2017).

15. Yoshikawa, N. *et al.* Interband resonant high-harmonic generation by valley polarized electron-hole pairs. *Nat. Commun.* **10**, 1–7 (2019).

16. Zaks, B., Liu, R. B. & Sherwin, M. S. Experimental observation of electron–hole recollisions. *Nature* **483**, 580–583 (2012).

17. Langer, F. *et al.* Lightwave-driven quasiparticle collisions on a subcycle timescale. *Nature* **533**, 225–229 (2016).

18. Banks, H. B. *et al.* Dynamical birefringence: Electron-hole recollisions as probes of Berry curvature. *Phys. Rev. X* **7**, 041042 (2017).

19. Langer, F. *et al.* Lightwave valleytronics in a monolayer of tungsten diselenide. *Nature* **557**, 76–80 (2018).

20. Saito, N. *et al.* Observation of selection rules for circularly polarized fields in high-harmonic generation from a crystalline solid. *Optica* **4**, 1333–1336 (2017).

21. Morimoto, T., Po, H. C. & Vishwanath, A. Floquet topological phases protected by time glide symmetry. *Phys. Rev. B* **95**, 195155 (2017).

22. Xu, X., Yao, W., Xiao, D. & Heinz, T. F. Spin and pseudospins in layered transition metal dichalcogenides. *Nat. Phys.* **10**, 343–350 (2014).

23. Ghimire, S. *et al.* Generation and propagation of high-order harmonics in crystals. *Phys. Rev. A* **85**, 043836 (2012).

24. Vampa, G., You, Y. S., Liu, H., Ghimire, S. & Reis, D. A. Observation of backward high-harmonic emission from solids. *Opt. Express* **26**, 12210–12218 (2018).

25. Xia, P. *et al.* Nonlinear propagation effects in high harmonic generation in reflection and transmission from gallium arsenide. *Opt. Express* **26**, 29393–29400 (2018).

26. Konishi, K. *et al.* Polarization-controlled circular second-harmonic generation from metal hole arrays with threefold rotational symmetry. *Phys. Rev. Lett.* **112**, 135502 (2014).

27. Seyler, K. L. *et al.* Electrical control of second-harmonic generation in a $WSe_2$ monolayer transistor. *Nat. Nanotechnol.* **10**, 407–411 (2015).

28. Kojima, K. *et al.* Restoring the intrinsic optical properties of CVD-grown $MoS_2$ monolayers and their heterostructures. *Nanoscale* **11**, 12798–12803 (2019).

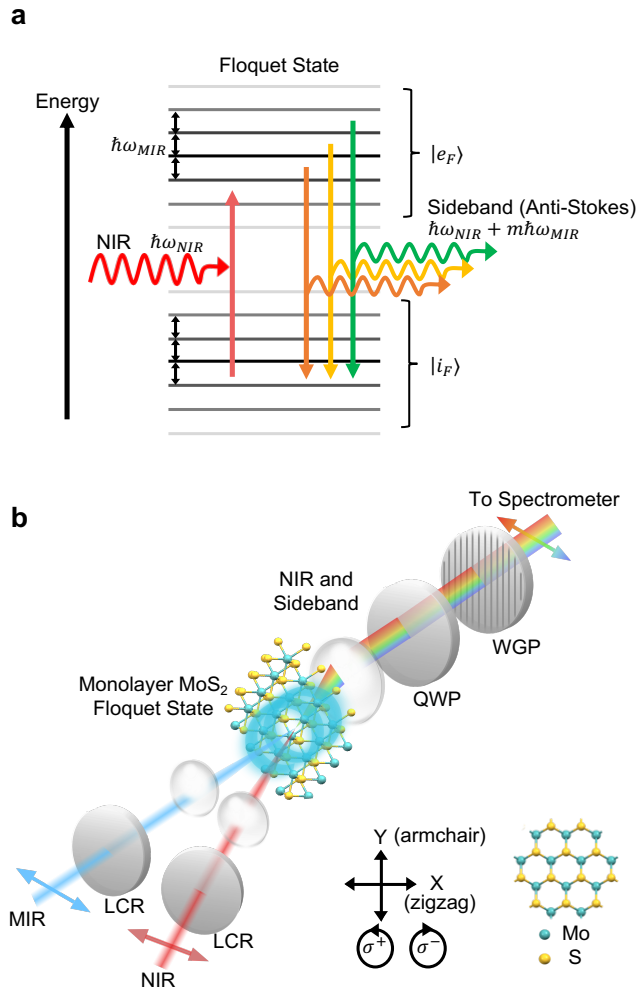

**Figure 1| Light scattering process in Floquet state.**

**a** Energy level diagram of light scattering process in MIR-driven Floquet state. The energies of the incident NIR photon and scattered *m*-th order sideband photon are written as $\hbar\omega_{NIR}$ and $\hbar\omega_{NIR} + m\hbar\omega_{MIR}$, respectively. The vertical arrows represent electronic transitions between the initial Floquet state $|i_F\rangle$ and intermediate Floquet state $|e_F\rangle$. Ladder-like levels with spacing of the MIR photon energy ($\hbar\omega_{MIR}$) denote the quasienergy levels of the Floquet states. **b** Schematic of HSG setup. (LCR: liquid crystal retarder, QWP: quarter wave plate, WGP: wire grid polarizer). The definition of the polarization is shown at the bottom right. The X and Y directions correspond to the zigzag and armchair directions of monolayer $MoS_2$

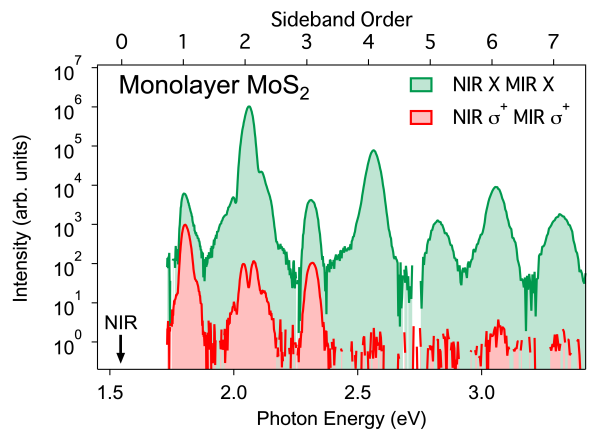

**Figure 2| HSG spectra from monolayer MoS$_2$.**

Polarization-unresolved HSG spectra measured under linearly polarized (green, X-polarized NIR and MIR pulses) and circularly polarized excitation (red, $\sigma^+$-polarized NIR and MIR pulses). The intensity of the green spectrum is multiplied by 100 for clarity. The black arrow shows the photon energy of the NIR pulses.

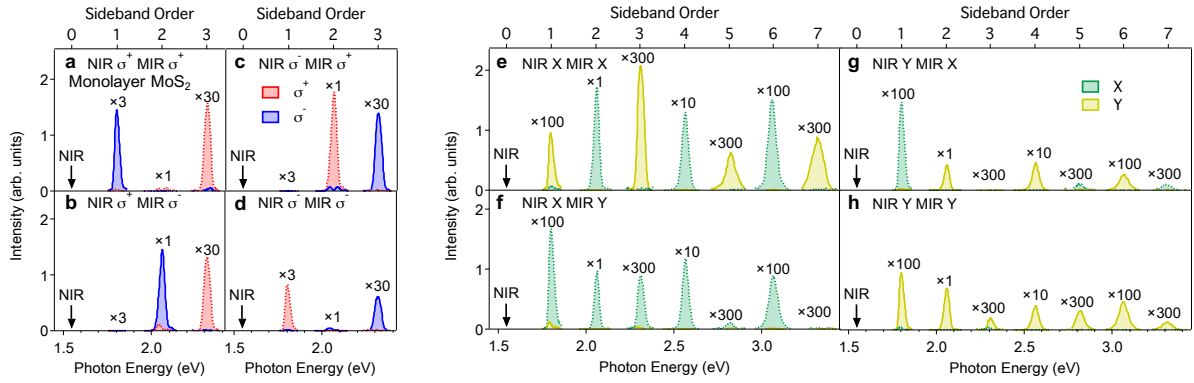

**Figure 3| Circular and linear polarization-resolved HSG spectra from monolayer MoS$_2$.**
**a-d** Circular polarization-resolved HSG spectra. Red and blue shaded areas indicate $\sigma^+$- and $\sigma^-$-polarized spectra, respectively. The sidebands are generated by **a** $\sigma^+$-polarized NIR and $\sigma^+$-polarized MIR pulses ($\sigma^+$, $\sigma^+$), **b** ($\sigma^+$, $\sigma^-$), **c** ($\sigma^-$, $\sigma^+$), **d** ($\sigma^-$, $\sigma^-$). **e-h** Linear polarization-resolved HSG spectra. Green and yellow areas indicate X- and Y-polarized spectra, respectively. The sidebands are generated by **e** (X, X), **f** (X, Y), **g** (Y, X), **h** (Y, Y). Each order of the sideband is scaled with the indicated number.

| NIR | $\sigma^+$ | $\sigma^+$ | $\sigma^-$ | $\sigma^-$ |
| --- | --- | --- | --- | --- |
| MIR | $\sigma^+$ | $\sigma^-$ | $\sigma^+$ | $\sigma^-$ |
| 1st | $\sigma^-$ | – | – | $\sigma^+$ |
| 2nd | – | $\sigma^-$ | $\sigma^+$ | – |
| 3rd | $\sigma^+$ | $\sigma^+$ | $\sigma^-$ | $\sigma^-$ |

| NIR | X | Y | X | Y |
| --- | --- | --- | --- | --- |
| MIR | X | X | Y | Y |
| Odd | Y | X | X | Y |
| Even | X | Y | X | Y |

**Table 1| Polarization selection rule of HSG in monolayer MoS$_2$.**
Allowed polarizations of the sideband are presented for each combination of NIR and MIR polarization. "-" indicates the forbidden sideband order.

# Supplementary Information

# Dynamical symmetry of strongly light-driven electronic system in crystalline solids

Kohei Nagai[1], Kento Uchida[1], Naotaka Yoshikawa[1], Takahiko Endo[2], Yasumitsu Miyata[2], & Koichiro Tanaka[1]

[1]Department of Physics, Graduate School of Science, Kyoto University, Sakyo-ku, Kyoto 606-8502, Japan.
[2]Department of Physics, Tokyo Metropolitan University, Hachioji, Tokyo 192-0397, Japan.

## 1. Extraction of HSG spectra

Figure S1 **a** shows total spectra of HSG and HHG (green) and HHG spectra (black) under linearly polarized excitation pulses observed from monolayer $MoS_2$. The HHG spectra were obtained in the condition that NIR pulses were 10s delayed from the MIR pulses. In our experiment, NIR photon energy is about 6 times larger than the MIR photon energy. Thus, we observed seventh order harmonics at the frequency close to that of the first order sideband. We obtained HSG spectra clearly by subtracting HHG spectra as explained in the main text. Figure S1 **b** shows the spectra under circularly polarized excitation pulses with the same excitation power. In this case, we did not observe HHG. This result is consistent with the previous paper[1].

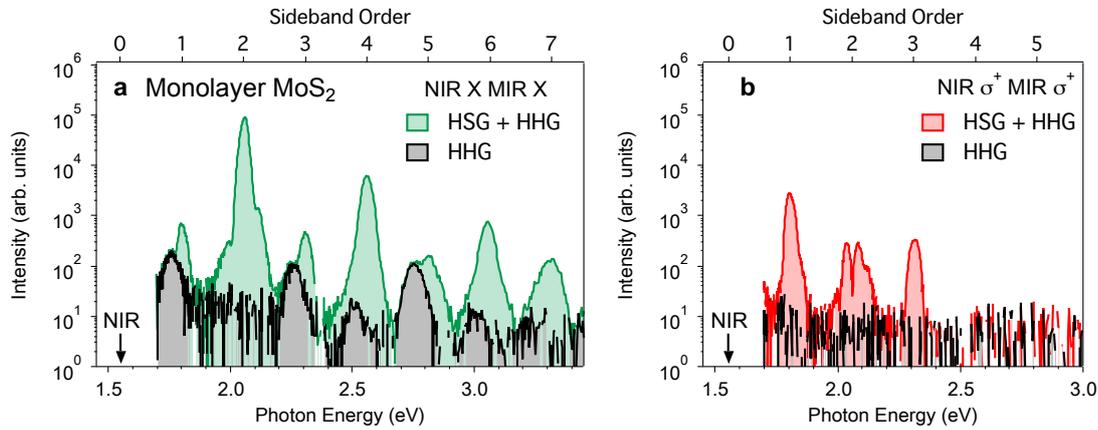

**Figure S1. Extraction of HSG spectra. a** Total spectra of HSG and HHG (color) and HHG spectra (black) under linear polarization excitation. **b** Same as **a** under circular polarization excitation.

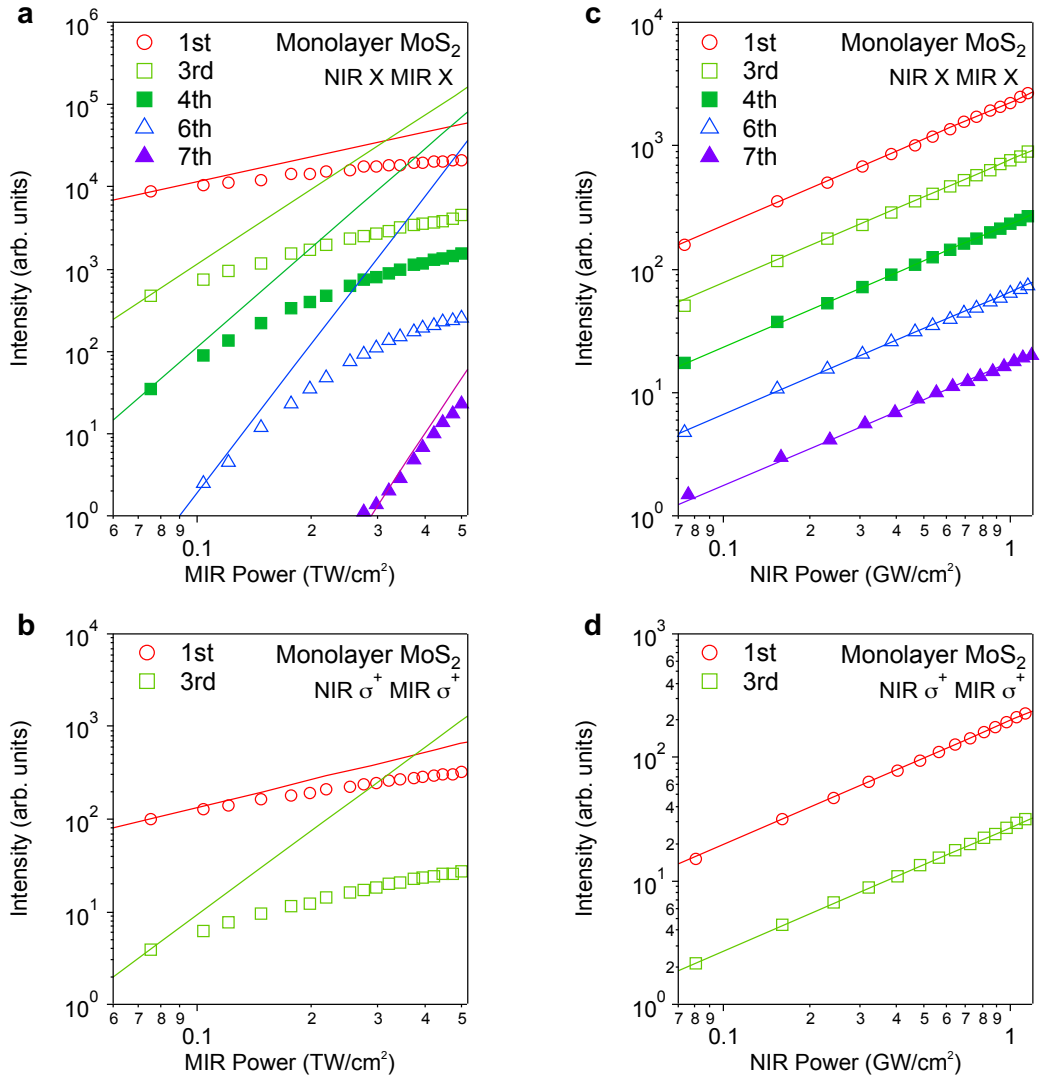

**Figure S2. Excitation power dependence of HSG in monolayer MoS$_2$. a,b** MIR power dependence of HSG under linear and circularly polarized light. **c,d** NIR power dependence of HSG under linearly and circularly polarized light. The power shown in the bottom axis is the peak power of the incident pulse at the focal point in vacuum. The solid lines are eye guides that show the power law.

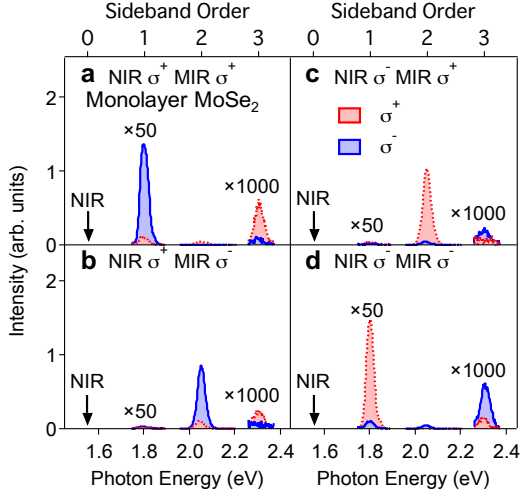

**Figure S3. Circular polarization-resolved HSG spectra from monolayer MoSe$_2$**

## 2. Microscopic description of HSG using Floquet theorem

In this section, we explain detailed descriptions of the microscopic model of HSG. We start with a single electron Hamiltonian with mass $m_e$ in a crystalline solid interacting with a MIR electric field given by,

$$\widehat{H}_0(\boldsymbol{r},t) = \frac{1}{2m_e}\left(\widehat{\boldsymbol{p}} - e\boldsymbol{A}(t)\right)^2 + U(\boldsymbol{r}), \tag{8}$$

where $e$ is the electron charge, $\boldsymbol{A}(t)$ is the vector potential of the external MIR field and $U(\widehat{\boldsymbol{r}})$ is periodic potential in the crystalline solid. Here the spatial distribution of $\boldsymbol{A}(t)$ is neglected by applying long wavelength approximation. Let us consider the additional weak NIR electric field, which perturbs the above system. The total Hamiltonian is given by

$$\widehat{H}(\boldsymbol{r},t) = \frac{1}{2m_e}\left(\widehat{\boldsymbol{p}} - e\boldsymbol{A}(t) - e\delta\boldsymbol{A}_{NIR}(t)\right)^2 + U(\boldsymbol{r}), \tag{9}$$

where $\delta\boldsymbol{A}_{NIR}(t)$ is the vector potential of the weak NIR field. By using Coulomb gauge $\nabla \cdot \delta\boldsymbol{A}_{NIR} = 0$ and neglecting the second order of $\delta\boldsymbol{A}_{NIR}(t)$, The total Hamiltonian is given by

$$\widehat{H}(\boldsymbol{r},t) = [\frac{1}{2m_e}\left(\widehat{\boldsymbol{p}} - e\boldsymbol{A}(t)\right)^2 + U(\boldsymbol{r}) + \frac{e}{m_e}\delta\boldsymbol{A}_{NIR}(t) \cdot \left(i\hbar\nabla + e\boldsymbol{A}(t)\right)]$$

$$= \widehat{H}_0(\boldsymbol{r},t) + \delta\boldsymbol{A}_{NIR}(t) \cdot \widehat{\boldsymbol{J}}, \tag{10}$$

where

$$\widehat{\boldsymbol{J}} = -\frac{e}{m_e}\left(\widehat{\boldsymbol{p}} - e\boldsymbol{A}(t)\right) = \frac{e}{m_e}\left(i\hbar\nabla + e\boldsymbol{A}(t)\right) = \frac{\partial\widehat{H}_0(\boldsymbol{r},t)}{\partial\boldsymbol{A}(t)} \tag{11}$$

is defined as the current operator in the velocity gauge. We regard $\delta H = \delta A_{NIR}(t) \cdot \hat{J}$ as the perturbation Hamiltonian.

First, we consider MIR-driven system $\hat{H}_0(\mathbf{r}, t)$. In our experimental condition, the MIR pulses are intense, coherent laser pulses and have several cycles within their pulse durations. Thus, to simplify the description of HSG, we assume the time periodicity of the MIR field and apply the Floquet theorem[2]. Note that the Floquet concept is valid in both perturbative and nonperturbative regime. With this assumption, the Hamiltonian $\hat{H}_0(\mathbf{r}, t')$ satisfies the periodicity of the MIR field

$$\hat{H}_0(\mathbf{r}, t) = \hat{H}_0(\mathbf{r}, t + 2\pi/\omega_{MIR}) \tag{12}$$

with angular frequency of the MIR light $\omega_{MIR}$. Schrödinger equation under this Hamiltonian is given by

$$\left(\hat{H}_0(\mathbf{r}, t) - i\hbar \frac{\partial}{\partial t}\right) |\Psi(t)\rangle = 0 \tag{13}$$

where $|\Psi(t)\rangle$ is a wave function. By substituting this wavefunction with

$$|\Psi_\alpha(t)\rangle = \exp(-i\epsilon_\alpha t/\hbar) |\Phi_\alpha(t)\rangle, \tag{14}$$

eq. (13) can be written as an eigenvalue problem:

$$\hat{H}_F |\Phi_\alpha(t)\rangle = \epsilon_\alpha |\Phi_\alpha(t)\rangle, \tag{15}$$

where

$$\hat{H}_F = \hat{H}_0(\mathbf{r}, t) - i\hbar \frac{\partial}{\partial t} \tag{16}$$

is the Floquet Hamiltonian, and

$$|\Phi_\alpha(t)\rangle = |\Phi_\alpha(t + 2\pi/\omega_{MIR})\rangle. \tag{17}$$

is a Floquet eigenstate. Since $\hat{H}_F$ is a Hermitian operator, the eigenvalue $\epsilon_\alpha$ in eq. (15) is a real number, which is defined as quasienergy. We assume that the electronic system is in a Floquet eigenstate $|i_F(\mathbf{r}, t)\rangle$ at $t = -\infty$.

Next we consider the temporal evolution of the total system perturbed by the NIR field. Schrödinger equation under the Hamiltonian $H(t)$ is given by

$$i\hbar \frac{\partial}{\partial t} |\Psi(\mathbf{r}, t)\rangle = \left(\hat{H}_0(\mathbf{r}, t') + \delta \hat{H}(\mathbf{r}, t)\right) |\Psi(\mathbf{r}, t)\rangle \tag{18}$$

We assume the following ansatz for the wave function:

$$|\Psi(\mathbf{r}, t)\rangle = e^{-i\frac{\epsilon_{i_F}}{\hbar}t} |i_F(\mathbf{r}, t)\rangle + \sum_{e_F \neq i_F} C_{e_F}(t) e^{-i\frac{\epsilon_{e_F}}{\hbar}t} |e_F(\mathbf{r}, t)\rangle \tag{19}$$

where $|e_F(\mathbf{r}, t)\rangle$ denotes other Floquet eigenstates than $|i_F(\mathbf{r}, t)\rangle$. Here, the coefficient $C_{e_F}(t)$ is small because the NIR light is weak. By considering the first order perturbation, we obtain

$$C_{e_F}(t) = -\frac{i}{\hbar} \int_{-\infty}^{t} dt' e^{i\omega_{ei}t'} \langle e_F(\mathbf{r}, t')| \frac{\partial \hat{H}_0(\mathbf{r}, t')}{\partial \mathbf{A}(t')} |i_F(\mathbf{r}, t')\rangle \delta A_{NIR}(t') \tag{20}$$

with $\hbar \omega_{ei} = \epsilon_{e_F} - \epsilon_{i_F}$.

We calculate HSG by considering current induced in the perturbed state $|\Psi(\mathbf{r},t)\rangle$. The expectation value of the current is given by

$$J(t) = \langle \Psi(\mathbf{r},t)|\hat{J}|\Psi(\mathbf{r},t)\rangle$$
$$\simeq \langle i_F(\mathbf{r},t)|\hat{J}|i_F(\mathbf{r},t)\rangle + \sum_{e_F \neq i_F} \left( C_{e_F}(t)e^{-i\omega_{ei}t}\langle i_F(\mathbf{r},t)|\hat{J}|e_F(\mathbf{r},t)\rangle + c.c. \right), \quad (21)$$

where the second order term of $C_{e_F}(t)$ is neglected. The first term, which is independent of the NIR light field, contributes to HHG as discussed in ref. 3. HSG is induced by the second term ($J_{SG}(t)$) that is proportional to the NIR light field. By substituting eq. (20) into eq. (21), polarization current $J_{SG}(t)$ is given by

$$J_{SG,\mu}(t) = -\frac{i}{\hbar}\sum_{\nu}\sum_{e_F \neq i_F}\int_{-\infty}^{t}dt'\, e^{-i\omega_{ei}(t-t')}\chi^{e_F}_{\mu,\nu}(t,t')\delta A_{NIR,\nu}(t') + c.c. \quad (22)$$

with

$$\chi^{e_F}_{\mu,\nu}(t,t') = \left\langle i_F(\mathbf{r},t)\left|\frac{\partial \hat{H}_0(\mathbf{r},t)}{\partial A_\mu(t)}\right|e_F(\mathbf{r},t)\right\rangle \left\langle e_F(\mathbf{r},t')\left|\frac{\partial \hat{H}_0(\mathbf{r},t')}{\partial A_\nu(t')}\right|i_F(\mathbf{r},t')\right\rangle. \quad (23)$$

The tensor $\chi^{e_F}_{\mu,\nu}(t,t')$ satisfies the following temporal periodicity:

$$\chi^{e_F}_{\mu,\nu}(t,t') = \chi^{e_F}_{\mu,\nu}(t+2\pi l/\omega_{MIR}, t'+2\pi l'/\omega_{MIR}), \quad (24)$$

where $l, l'$ are integers. Thus, $\chi^{e_F}_{\mu,\nu}(t,t')$ can be expanded in a Fourier series as follows:

$$\chi^{e_F}_{\mu,\nu}(t,t') = \sum_{l,l'} a^{e_F}_{\mu,\nu,l,l'}\, e^{il\omega_{MIR}t}e^{-il'\omega_{MIR}t'}. \quad (25)$$

When the NIR light is a continuous wave, the Fourier component of $J_{SG}(t)$ can be written explicitly. We write the NIR light as $\delta A_{NIR,\nu}(t') = \delta A_{0,\nu}e^{-i\omega_{NIR}t'}$ by applying the rotating wave approximation. To simplify the discussion, we only consider the first term of eq.(22) because the complex conjugate term just gives the time-reversal pair of the current. We consider the damping $\Gamma$ and obtain

$$J_{SG,\mu}(\omega) = -\frac{i}{\hbar}\sum_{\nu,l,l'}\sum_{e_F \neq i_F} a^{e_F}_{\mu,\nu,l,l'}\delta A_{0,\nu}\int_{-\infty}^{\infty}dt\int_{-\infty}^{t}dt'\, e^{-(i\omega_{ei}+\Gamma)(t-t')}$$
$$\times e^{il\omega_{MIR}t}e^{-il'\omega_{MIR}t'}e^{-i\omega_{NIR}t'}e^{i\omega t}. \quad (26)$$

By calculating the integral with respect to $t'$ and taking the limit of $\Gamma \to +0$, we get

$$J_{SG,\mu}(\omega) = -\frac{i}{\hbar}\sum_{\nu,l,l'}\sum_{e_F \neq i_F} a^{e_F}_{\mu,\nu,l,l'}\delta A_{0,\nu}\int_{-\infty}^{\infty}dt\,\frac{e^{i(l-l')\omega_{MIR}-\omega_{NIR}+\omega)t}}{-l'\omega_{MIR}-\omega_{NIR}+\omega_{ei}-i0}$$
$$= -\frac{2\pi i}{\hbar}\sum_{\nu,l,l'}\sum_{e_F \neq i_F} a^{e_F}_{\mu,\nu,l,l'}\delta A_{0,\nu}\frac{1}{-l'\omega_{MIR}-\omega_{NIR}+\omega_{ei}-i0}$$

$$\times \delta((l - l')\omega_{MIR} - \omega_{NIR} + \omega)) \tag{27}$$

Hence, the current $\mathbf{J}_{SG}(t)$ is the sum of the $m$-th order frequency component written as

$$J_{SG,\mu}(t) = \sum_m J_{SG,m,\mu}(t)$$

$$= -\frac{i}{\hbar} \sum_m \sum_{v,l} \sum_{e_F \neq i_F} a^{e_F}_{\mu,v,l,l'} \delta A_{0,v} \frac{1}{-(m+l)\omega_{MIR} - \omega_{NIR} + \omega_{ei} - i0} e^{-i(\omega_{NIR} + m\omega_{MIR})t}. \tag{28}$$

The current $\mathbf{J}_{SG}(t)$ emits the sideband light according to Maxwell equation. If we consider a simple case where current uniformly in xy plane at z = 0, the electric field induced by the current is given by

$$\mathbf{E}_{SG,m}(z,t) = \mu_0 c \mathbf{J}_{SG,m}\left(t - \frac{z}{c}\right) \tag{29}$$

where $\mu_0$ is magnetic permeability in a vacuum and $c$ is the speed of light.

## 3. Derivation of symmetry constraint on "Raman tensor"

In this section, we derive the symmetry constraint on "Raman tensor" discussed in the main text.

**Definition of dynamical symmetry**

Here, we show the definition of the DS operation used in this paper. We basically follow the definition and the notation of DS operations in ref. 4. Accordingly, time-reversal operation is included in our discussion of DS. We regard a "static" symmetry operation as one of the DS operations[5]. Note that we distinguish the following two different definitions of the DS operations.

1. Operations for the vector component and temporal part of the electric fields (denoted as $\hat{X}$)
2. Corresponding unitary and anti-unitary operations in quantum mechanics for the position of the electron and time $(\mathbf{r}, t)$ (denoted as $\hat{X}_q$)

**Symmetry constraint on microscopic model**

We derive the DS constraint on the microscopic model introduced in the section 2. We show only the example for the DS operations that are accompanied by the temporal translation. Note that the microscopic model is also restricted by the DS operations accompanied by the time-reversal operation.

In the following, to simplify the discussion, we only consider the first term of eq.(22). We consider the DS operation $\hat{X}^t = \hat{X}_s \cdot \hat{\tau}^t$ for the MIR driven system, where $\hat{X}_s$ is an operation for the spatial component of the vectors and $\hat{\tau}^t$ is a time-translation operation for the time $t$ by $\Delta t$. By applying this operation for the current $\mathbf{J}_{SG}(t)$ in eq. (22), we obtain

$$\hat{X}^t J_{SG}(t) = -\frac{i}{\hbar} \sum_{e_F \neq i_F} \hat{X}^t \int_{-\infty}^{t} dt' e^{-i\omega_{ei}(t-t')} \chi^{e_F}(t,t') \delta A_{NIR}(t')$$

$$= -\frac{i}{\hbar} \sum_{e_F \neq i_F} \int_{-\infty}^{t+\Delta t} dt' e^{-i\omega_{ei}(t+\Delta t-t')} \hat{X}_s \chi^{e_F}(t+\Delta t, t') \hat{X}_s^{-1} \hat{X}_s \delta A_{NIR}(t')$$

$$= -\frac{i}{\hbar} \sum_{e_F \neq i_F} \int_{-\infty}^{t} dt'' e^{-i\omega_{ei}(t-t'')} \hat{X}_s \chi^{e_F}(t+\Delta t, t''+\Delta t) \hat{X}_s^{-1} \hat{X}_s \delta A_{NIR}(t''+\Delta t)$$

$$= -\frac{i}{\hbar} \sum_{e_F \neq i_F} \int_{-\infty}^{t} dt'' e^{-i\omega_{ei}(t-t'')} \hat{X}^t \chi^{e_F}(t, t'') (\hat{X}^{t''})^{-1} \hat{X}^{t''} \delta A_{NIR}(t''),$$

(30)

where $\hat{X}^{t''}$ operates on time $t''$.

In the following, we demonstrate the equation

$$\hat{X}^t \chi^{e_F}(t,t'') (\hat{X}^{t''})^{-1} = \chi^{e_F}(t,t''). \tag{31}$$

For preparation, we discuss the symmetry of the Hamiltonian and wave functions. When $\hat{X}^t$ is a DS operator of the MIR driven system, the corresponding operator commutes with the Hamiltonian (eq. (8)) as following:

$$\hat{X}_q^t \hat{H}_0(\mathbf{r}, t) = \hat{H}_0(\mathbf{r}, t) \hat{X}_q^t. \tag{32}$$

The same should hold for the Floquet Hamiltonian

$$\hat{X}_q^t \hat{H}_F(\mathbf{r}, t) = \hat{H}_F(\mathbf{r}, t) \hat{X}_q^t. \tag{33}$$

We consider the simplest case where the eigenstates of $\hat{H}_F(\mathbf{r}, t)$ are nondegenerate. In this case, the eigenstates are simultaneous eigenstates of $\hat{H}_F(\mathbf{r}, t)$ and $\hat{X}^t$. Hence, for eigenstate $|\Phi_F(\mathbf{r}, t)\rangle$ of $\hat{H}_F(\mathbf{r}, t)$,

$$\hat{X}_q^t |\Phi_F(\mathbf{r}, t)\rangle = e^{i\delta} |\Phi_F(\mathbf{r}, t)\rangle \tag{34}$$

holds, where the phase $\delta$ is a real number. As a result, we get

$$\hat{X}^t \left\langle i_F(\mathbf{r}, t) \left| \frac{\partial \hat{H}_0(\mathbf{r},t)}{\partial A(t)} \right| e_F(\mathbf{r}, t) \right\rangle \equiv \left\langle i_F(\mathbf{r}, t) \left| (\hat{X}_q^t)^\dagger \frac{\partial \hat{H}_0(\mathbf{r},t)}{\partial A(t)} \hat{X}_q^t \right| e_F(\mathbf{r}, t) \right\rangle$$

$$= \left\langle i_F(\mathbf{r}, t) \left| \frac{\partial \hat{H}_0(\mathbf{r},t)}{\partial A(t)} \right| e_F(\mathbf{r}, t) \right\rangle e^{i(\delta_i - \delta_e)}, \tag{35}$$

where $\delta_i$ and $\delta_e$ are the phases in eq (34) corresponding to $|i_F\rangle$ and $|e_F\rangle$ respectively. Through the same discussion for $\hat{X}^{t''}$, eq. (31) can be demonstrated.

By combining eq. (31) with eq. (30), we obtain

$$\hat{X}^t \boldsymbol{J}_{SG}(t) = -\frac{i}{\hbar} \sum_{e_F \neq i_F} \int_{-\infty}^{t} dt'' e^{-i\omega_{ei}(t-t'')} \chi^{e_F}(t,t'') \hat{X}^{t''} \delta \boldsymbol{A}_{NIR}(t'') + c.c.. \tag{36}$$

This equation gives a relation between the polarization of the NIR and the sideband. Note that this discussion is valid even in the case that the eigenstates are degenerate through the same discussion of the degeneracy in ref 4.

**Derivation of symmetry constraint on "Raman tensor"**

The relation in eq. (36) can be transformed into a simpler form as the invariance of "Raman tensor" under DS operation. When the NIR light is a continuous wave, any vector potential $\delta \boldsymbol{A}_{NIR}(t'')$ can be written as a linear combination of "eigenvectors" of DS operator, which are summarized in Table S1. For example, if $\hat{C}_{3,\sigma_M}$, which is shown in the main text, is a DS operator of the system, $\delta \boldsymbol{A}_{NIR}(t) = \delta A_{NIR}(1,-i)^T e^{i\omega_{NIR}t}$ is an eigenvector satisfying

$$\hat{C}_{3,\sigma_M} \begin{pmatrix} 1 \\ -i \end{pmatrix} e^{i\omega_{NIR}t} = e^{i\frac{2\pi}{3}} e^{-i\frac{2\pi\omega_{NIR}}{3\omega_{MIR}}\sigma_M} \begin{pmatrix} 1 \\ -i \end{pmatrix} e^{i\omega_{NIR}t}, \tag{37}$$

where $\omega_{NIR}$ is angular frequency of the NIR light. When $\delta \boldsymbol{A}_{NIR}(t)$ satisfies
$$\hat{X}^t \delta \boldsymbol{A}_{NIR}(t) = e^{i\theta} \delta \boldsymbol{A}_{NIR}(t),$$
the tensor product of $\hat{X}^t \boldsymbol{J}_{SG}(t)$ and $\hat{X}^t \delta \boldsymbol{A}_{NIR}(t)$ is given by

$$\hat{X}^t \boldsymbol{J}_{SG}(t) \left( \hat{X}^t \boldsymbol{A}_{NIR}(t) \right)^\dagger$$

$$= -\frac{i}{\hbar} \sum_{e_F \neq i_F} \int_{-\infty}^{t} dt'' e^{-i\omega_{ei}(t-t'')} \chi^{e_F}(t,t') \hat{X}^{t''} \delta \boldsymbol{A}_{NIR}(t'') \left( \hat{X}^t \boldsymbol{A}_{NIR}(t) \right)^\dagger$$

$$= -\frac{i}{\hbar} \sum_{e_F \neq i_F} \int_{-\infty}^{t} dt'' e^{-i\omega_{ei}(t-t'')} \chi^{e_F}(t,t') \delta \boldsymbol{A}_{NIR}(t'') e^{i\theta} \delta \boldsymbol{A}_{NIR}^\dagger(t) e^{-i\theta}$$

$$= -\frac{i}{\hbar} \sum_{e_F \neq i_F} \int_{-\infty}^{t} dt'' e^{-i\omega_{ei}(t-t'')} \chi^{e_F}(t,t') \delta \boldsymbol{A}_{NIR}(t'') \delta \boldsymbol{A}_{NIR}^\dagger(t)$$

$$= \boldsymbol{J}_{SG}(t) \delta \boldsymbol{A}_{NIR}^\dagger(t). \tag{38}$$

According to eq.(28), the symmetry constraint can be reduced into

$$\hat{X}^t \boldsymbol{J}_{SG,m}(t) \left( \hat{X}^t \delta \boldsymbol{A}_{NIR}(t) \right)^\dagger = \boldsymbol{J}_{SG,m}(t) \delta \boldsymbol{A}_{NIR}^\dagger(t) \tag{39}$$

by comparing the $m$-th order frequency component of both sides in eq. (39). When the electric field of the sideband is given by eq. (28), by considering $E_{NIR}(t) = -\frac{\partial(\delta A_{NIR}(t))}{\partial t}$, we get

$$\hat{X}^t E_{SG,m}(t) \left(\hat{X}^t E_{NIR}(t)\right)^\dagger = E_{SG,m}(t) E_{NIR}(t)^\dagger \tag{40}$$

This equation indicates the invariance of "Raman tensor" under the DS operation :

$$\hat{X}^t \mathcal{R}_m(t) = \mathcal{R}_m(t) \tag{41}$$

$$\mathcal{R}_m(t) = E_{SG,m}(t) E_{NIR}^\dagger(t). \tag{42}$$

**Table S1| Dynamical symmetry operation and eigenvectors**

| Dynamical symmetry operation | | Eigenvector | | Eigenvalue |
|---|---|---|---|---|
| Spatial operation | $\hat{\sigma}_y$ | Spatial part | $\begin{pmatrix}1\\0\end{pmatrix}, \begin{pmatrix}0\\1\end{pmatrix}$ | $+1, -1$ |
| | $\hat{R}_n$ | | $\begin{pmatrix}1\\-i\end{pmatrix}, \begin{pmatrix}1\\i\end{pmatrix}$ | $e^{i\frac{2\pi}{n}}, e^{-i\frac{2\pi}{n}}$ |
| Temporal operation | $\hat{\tau}_n$ | Temporal part (NIR) | $e^{i\omega_{NIR} t}$ | $e^{-i\frac{2\pi \omega_{NIR}}{n \omega_{MIR}}}$ |
| | $\hat{T}$ | | $\cos(\omega_{NIR} t), \sin(\omega_{NIR} t)$ | $+1, -1$ |
| | $\hat{T} \cdot \hat{\tau}_2$ | | $\cos\left(\omega_{NIR} t - \frac{\omega_{NIR} \pi}{2\omega_{MIR}}\right),$ $\sin\left(\omega_{NIR} t - \frac{\omega_{NIR} \pi}{2\omega_{MIR}}\right)$ | $+1, -1$ |

## 4. Determination of dynamical symmetry in the Floquet systems

In this section, DSs under linearly and circularly polarized light in monolayer TMDs are derived. To derive all selection rules, all symmetry operations must be considered. The DS operations of the MIR-driven Floquet system in solids are derived as the following procedure:
1. Identify the symmetry of crystal only
2. Identify the symmetry of MIR light only
3. Identify the intersection of the DS groups of the crystal and MIR field

### 4.1 Crystal symmetry of monolayer TMDs

The symmetry operations of crystalline solids are well classified by the crystallographic point group. Monolayer $MoS_2$ and $MoSe_2$ belongs to $D_{3h}$ point group. In our experimental condition, the symmetry in the 2D space parallel to the sample is enough to describe the selection rules. The point group to describe the symmetry in the 2D space are generated by two operations $\hat{R}_3$ and $\hat{\sigma}_v$, where $\hat{R}_3$ is a rotation by $2\pi/3$, $\hat{\sigma}_v$ is a mirror operation with respect to the Mo-S direction in the crystal. Since the symmetry of the crystal is time-independent, the DS groups of the monolayers are generated by $\hat{R}_3$, $\hat{\sigma}_v$, infinitesimal temporal translation, and time-reversal operation.

### 4.2 Dynamical symmetry of MIR light

We describe DS in the product space of time and the 2D space where the polarization is defined. In the following, we determine the DS of the MIR field by checking the invariance under all DS operations given in ref 4.

**Case 1| Circularly polarized MIR light**

We write the electric field of circularly polarized light as follows:

$$\boldsymbol{E}(t) = \begin{pmatrix} \sin(\omega_{MIR} t) \\ -\sigma_M \cos(\omega_{MIR} t) \end{pmatrix}. \tag{43}$$

In this representation, DS group of the MIR light is generated by

$$\widehat{D}_y = \widehat{T} \cdot \hat{\sigma}_y \tag{44}$$

$$\widehat{C}_{n,\sigma_M} = \hat{\tau}_{-n\sigma_M} \cdot \widehat{R}_n \ (n \to \infty), \tag{45}$$

where $\widehat{T}$ is time-reversal symmetry operation, $\hat{\sigma}_y$ is mirror symmetry operation with respect to y axis, $\widehat{R}_n$ is the spatial rotation by $2\pi/n$, $\hat{\tau}_n$ is the temporal translation by $T/n$ ($T = 2\pi/\omega_{MIR}$), and $\sigma_M = \pm 1$ represent left and right circular polarization of the MIR light respectively. $\widehat{C}_{n,\sigma_M}$ ($n \to \infty$) denotes a DS operation with the infinitesimal spatial rotation and infinitesimal temporal translation.

**Case 2| Linear polarized MIR light**

We write the electric field of x- polarized light as follows:

$$\boldsymbol{E}(t) = \begin{pmatrix} 1 \\ 0 \end{pmatrix} \sin(\omega_{MIR} t). \tag{46}$$

In this representation, DS group of the MIR light is generated by the following operations

$$\hat{\sigma}_x \tag{47}$$
$$\widehat{C}_2 \tag{48}$$
$$\widehat{D}_y \tag{49}$$
$$\widehat{Z}_y = \hat{\tau}_2 \cdot \hat{\sigma}_y \tag{50}$$
$$\widehat{H}_x = \widehat{T} \cdot \hat{\tau}_2 \cdot \hat{\sigma}_x \tag{51}$$
$$\widehat{Q} = \widehat{T} \cdot \widehat{R}_2 \tag{52}$$

Similarly, the DS group of the y- polarized light MIR light can be written by substituting $x$ and $y$.

**4.3 Dynamical symmetry of MIR driven Floquet state in monolayer TMDs**

The DS group of the MIR-driven Floquet state in monolayer TMDs are the intersection of the DS groups of the crystal and MIR field. By comparing the DS groups discussed in the section 4.1 and 4.2, the generators of the DS group of the MIR-driven Floquet state are obtained as follows:

**Case 1| Circularly polarized MIR light**

$$\hat{C}_{3,\sigma_M}, \hat{D}_y \tag{53}$$

**Case 2| x- polarized MIR light**

$$\hat{D}_y, \hat{Z}_y \tag{54}$$

**Case 3| y- polarized MIR light**

$$\hat{\sigma}_y, \hat{H}_y \tag{55}$$

## 5. Dynamical symmetry on "Raman tensor" and selection rules

We require the invariance of "Raman tensor" under DS operations to derive selection rules. In the following, we show two examples of the derivations of the selection rules. All symmetry constraints under the circularly and linearly polarized MIR field are summarized in Table S2.

Table S2| **DS operation and selection rule of HSG**

| Dynamical symmetry operation | Selection rule of HSG |
|---|---|
| $\hat{D}_y$ | x or y- polarized NIR light <br> ↓ <br> Elliptically polarized sideband light with major or minor axis parallel to the x-axis |
| $\hat{C}_{n,\sigma_M}$ | Circularly polarized NIR light with helicity $\sigma_N$ <br> ↓ <br> Circularly polarized $(\sigma_M(-\sigma_N \pm 1) + nN)$ –th (N : integer) order sideband light with helicity ($\pm 1$) |
| $\hat{Z}_y$ | x or y- polarized NIR light <br> ↓ <br> Odd order sideband light polarized perpendicularly to NIR light <br> Even order sideband light polarized in parallel with NIR light |
| $\hat{\sigma}_y$ | x or y- polarized NIR light <br> ↓ <br> Sideband light polarized in parallel with NIR light |
| $\hat{H}_y$ | x or y- polarized NIR light <br> ↓ <br> Elliptically polarized sideband light with major or minor axis parallel to the x-axis |

**Case 1| Circularly polarized MIR light**

**Invariance under $\widehat{D}_y$**

We write the "Raman tensor" by using an electric field of the NIR and $m$-th order sideband light as follows:

$$\mathcal{R}_m(t) = \boldsymbol{E}_{SG,m}(t)\boldsymbol{E}_{NIR}^\dagger(t)$$

$$= \sin\bigl((\omega_{NIR} + m\omega_{MIR})t + \phi_S\bigr)\sin(\omega_{NIR}t + \phi_N)\begin{pmatrix} E_{S,m,x}E_{NIR,x}^* & E_{S,m,x}E_{NIR,y}^* \\ E_{S,m,y}E_{NIR,x}^* & E_{S,m,y}E_{NIR,y}^* \end{pmatrix}, \quad (56)$$

where $\boldsymbol{E}_{SG,m}(t) = \sin((\omega_{NIR} + m\omega_{MIR})t + \phi_S)\bigl(E_{SG,m,x}, E_{SG,m,y}\bigr)^T$ and $\boldsymbol{E}_{NIR}(t) = \sin(\omega_{NIR}t + \phi_N)\bigl(E_{NIR,x}, E_{NIR,y}\bigr)^T$ denote the electric field of the sideband and NIR light respectively. Note that $\bigl(E_{NIR,x}, E_{NIR,y}\bigr)^T$ must be $(1,0)^T$ or $(0,1)^T$, and $\phi_N$ must be 0 or $\pi/2$ under $\widehat{D}_y$ because the electric field of the NIR and sideband light must be eigenvectors of $\widehat{D}_y$. $\phi_N$ can be set to 0 without loss of generality. By requiring $\widehat{D}_y \mathcal{R}_m = \mathcal{R}_m$, the relation between the electric field of the NIR and sideband light can be obtained.

For example, under y-polarized NIR field (i.e., $E_{NIR,x} = 0$), the electric field of the sideband light is constrained as

$$E_{SG,m,x} = 0 \text{ and } \phi_S = 0 \quad (57)$$

or

$$E_{SG,m,y} = 0 \text{ and } \phi_S = \pi/2. \quad (58)$$

The sideband can be a linear combination of allowed two cases:

$$\boldsymbol{E}_{SG,m}(t) = \begin{pmatrix} \cos\bigl((\omega_{NIR} + m\omega_{MIR})t\bigr) \\ b\sin\bigl((\omega_{NIR} + m\omega_{MIR})t\bigr) \end{pmatrix}, \quad (59)$$

where $b$ is a real number. This demonstrates that the sideband light can be elliptical polarization with major or minor axis parallel to the x-axis. In the same way, constraints on the sideband light can also be determined under x-polarized NIR field. Thus, the selection rule for $\widehat{D}_y$ in Table S2 is obtained, which is similar to the result of the selection rules of HHG[4].

**Invariance under $\widehat{C}_{n,\sigma_M}$**

Here, we discuss the invariance of "Raman tensor" under $\widehat{C}_{n,\sigma_M}$ ($n$ : integer). We write the "Raman tensor" by using the electric field of the NIR and $m$-th order sideband light as follows:

$$\mathcal{R}_m(t) = \boldsymbol{E}_{S,m}(t)\boldsymbol{E}_{NIR}^\dagger(t) = \exp(im\omega_{MIR}t)\begin{pmatrix} 1 & i\sigma_N \\ -i\sigma_S^m & \sigma_S^m\sigma_N \end{pmatrix}, \quad (60)$$

where we take circular polarization bases: $E_{S,m} = \exp(i(\omega_{NIR} + m\omega_{MIR})t)(1, -i\sigma_S^m)^T$, $E_{NIR} = \exp(i\omega_{NIR}t)(1, -i\sigma_N)^T$. By requiring $\hat{C}_{n,\sigma_M}\mathcal{R}_m = \mathcal{R}_m$, we obtain

$$(\hat{\tau}_{-n\sigma_M}\exp(im\omega_{MIR}t))\left(\hat{R}_n\begin{pmatrix}1 & i\sigma_N \\ -i\sigma_S^m & \sigma_N\sigma_S^m\end{pmatrix}\right)$$

$$= \exp(im\omega_{MIR}t)\begin{pmatrix}1 & i\sigma_N \\ -i\sigma_S^m & \sigma_N\sigma_S^m\end{pmatrix} \tag{61}$$

$$\exp\left(i\left(m\omega_{MIR}t - m\frac{2\pi\sigma_M}{n}\right)\right)\exp\left(i(\sigma_S^m - \sigma_N)\frac{2\pi}{n}\right)\begin{pmatrix}1 & i\sigma_N \\ -i\sigma_S^m & \sigma_N\sigma_S^m\end{pmatrix}$$

$$= \exp(im\omega_{MIR}t)\begin{pmatrix}1 & i\sigma_N \\ -i\sigma_S^m & \sigma_N\sigma_S^m\end{pmatrix}, \tag{62}$$

where the operation of the rotation $\hat{R}_n$ to the matrix part is represented by

$$\hat{R}_n\begin{pmatrix}1 & i\sigma_N \\ -i\sigma_S^m & \sigma_N\sigma_S^m\end{pmatrix}$$

$$= \begin{pmatrix}\cos(2\pi/n) & -\sin(2\pi/n) \\ \sin(2\pi/n) & \cos(2\pi/n)\end{pmatrix}\begin{pmatrix}1 & i\sigma_N \\ -i\sigma_S^m & \sigma_N\sigma_S^m\end{pmatrix}\begin{pmatrix}\cos(2\pi/n) & \sin(2\pi/n) \\ -\sin(2\pi/n) & \cos(2\pi/n)\end{pmatrix}. \tag{63}$$

Thus, symmetry constraint can be written by

$$m\sigma_M + \sigma_N - \sigma_S^m = nN. \tag{64}$$

where $N$ is an integer. Note that $\hat{D}_y$ gives no restrictions for the sideband under circularly polarized NIR field.

We show an example of the circular polarization selection rules given in Table 1 in the main text. Since monolayer TMDs have threefold rotational symmetry, $n$ is 3 in our experiment. When both the MIR and NIR light is left circular polarized ($\sigma_M, \sigma_N = 1$), $\sigma_S^m$ must be equals to $m + 1 - 3N$ according to the eq. (64). Thus, $\sigma_S^1 = -1$ with $N = 1$, and $\sigma_S^3 = 1$ with $N = 0$ are allowed, but there is no solution for $\sigma_S^2$, which indicates the second order sideband is forbidden. The other cases of the circular polarization selection rules shown in Table 1 is also derived similarly.

**Case 2| x- polarized MIR light**
We have discussed the operation $\hat{D}_y$, thus, we only deal with the $\hat{Z}_y$.
**Invariance under $\hat{Z}_y$**
We write the "Raman tensor" by using the electric field of the NIR and $m$-th order sideband light as follows:

$$\mathcal{R}_m(t) = \boldsymbol{E}_{S,m}(t)\boldsymbol{E}_{NIR}^\dagger(t) = \exp(im\omega_{MIR}t)\begin{pmatrix}E_{S,m,x}E_{NIR,x}^* & E_{S,m,x}E_{NIR,y}^* \\ E_{S,m,y}E_{NIR,x}^* & E_{S,m,y}E_{NIR,y}^*\end{pmatrix}, \tag{65}$$

where $(E_{NIR,x}, E_{NIR,y})^T$ is $(1,0)^T$ or $(0,1)^T$. By requiring $\hat{Z}_y \mathcal{R}_m = \mathcal{R}_m$, we obtain

$$(-1)^m \begin{pmatrix} E_{S,m,x}E^*_{NIR,x} & E_{S,m,x}E^*_{NIR,y} \\ E_{S,m,y}E^*_{NIR,x} & E_{S,m,y}E^*_{NIR,y} \end{pmatrix} = \begin{pmatrix} E_{S,m,x}E^*_{NIR,x} & -E_{S,m,x}E^*_{NIR,y} \\ -E_{S,m,y}E^*_{NIR,x} & E_{S,m,y}E^*_{NIR,y} \end{pmatrix}, \tag{66}$$

Here, the term $(-1)^m$ is attributed to the half-period temporal translation in $\hat{Z}_y$. The electric field of the sideband light is constrained as

$$E_{S,m,y} = 0 \ (m: \text{even}) \text{ and } E_{S,m,x} = 0 \ (m: \text{odd}) \tag{67}$$

under y-polarized NIR field (i.e., $E_{NIR,x} = 0$) and

$$E_{S,m,x} = 0 \ (m: \text{even}) \text{ and } E_{S,m,y} = 0 \ (m: \text{odd}) \tag{68}$$

under x-polarized NIR field (i.e., $E_{NIR,y} = 0$). As a result, the polarizations of the odd-order sidebands are perpendicular to the NIR field and even-order sidebands are parallel. Note that this selection rule is stricter than that derived from $\hat{D}_y$.

We also note that Langer *et al.* observed elliptically polarized sidebands in monolayer WSe$_2$ with a circularly polarized NIR field. This experimental result is consistent with our theory because both $\hat{D}_y$ and $\hat{Z}_y$ do not restrict the polarization of the sideband under NIR field.

**Case 3| y- polarized MIR light**
**Invariance under $\hat{\sigma}_y$**
We write the "Raman tensor" by using the electric field of the NIR and $m$-th order sideband light as follows:

$$\mathcal{R}_m(t) = \boldsymbol{E}_{S,m}(t)\boldsymbol{E}^\dagger_{NIR}(t) = \exp(im\omega_{MIR}t)\begin{pmatrix} E_{S,m,x}E^*_{NIR,x} & E_{S,m,x}E^*_{NIR,y} \\ E_{S,m,y}E^*_{NIR,x} & E_{S,m,y}E^*_{NIR,y} \end{pmatrix}, \tag{69}$$

where $(E_{NIR,x}, E_{NIR,y})^T$ is $(1,0)^T$ or $(0,1)^T$. By requiring $\hat{\sigma}_y \mathcal{R}_m = \mathcal{R}_m$, we obtain

$$E_{S,m,y} = 0 \tag{70}$$

under y-polarized NIR field (i.e., $E_{NIR,x} = 0$) and

$$E_{S,m,x} = 0 \tag{71}$$

under x-polarized NIR field (i.e., $E_{NIR,y} = 0$). This indicates that all sidebands have the same polarization as that of the NIR field.

**Invariance under $\hat{H}_y$**
We write the "Raman tensor" by using an electric field of the NIR and $m$-th order sideband light as follows:
$$\mathcal{R}_m(t) = \boldsymbol{E}_{SG,m}(t)\boldsymbol{E}^\dagger_{NIR}(t)$$

$$= \sin\left((\omega_{NIR} + m\omega_{MIR})t - \frac{(\omega_{NIR} + m\omega_{MIR})\pi}{2(\omega_{NIR} + m\omega_{MIR})} + \phi_S\right) \sin\left(\omega_{NIR}t - \frac{\omega_{NIR}\pi}{2\omega_{MIR}} + \phi_N\right)$$

$$\times \begin{pmatrix} E_{S,m,x}E_{NIR,x}^* & E_{S,m,x}E_{NIR,y}^* \\ E_{S,m,y}E_{NIR,x}^* & E_{S,m,y}E_{NIR,y}^* \end{pmatrix}, \tag{72}$$

where $(E_{NIR,x}, E_{NIR,y})^T$ is $(1,0)^T$ or $(0,1)^T$, and $\phi_N$ must be 0 or $\pi/2$ under $\widehat{D}_y$ because the electric field of the NIR and sideband light must be eigenvectors of $\widehat{D}_y$. $\phi_N$ can be set to 0 without loss of generality. Under y-polarized NIR field (i.e., $E_{NIR,x} = 0$), by requiring $\widehat{D}_y \mathcal{R}_m = \mathcal{R}_m$, the electric field of the sideband light is constrained as

$$E_{SG,m,x} = 0 \text{ and } \phi_S = 0 \tag{73}$$

or

$$E_{SG,m,y} = 0 \text{ and } \phi_S = \pi/2. \tag{74}$$

The sideband can be a linear combination of allowed two cases:

$$\boldsymbol{E}_{SG,m}(t) = \begin{pmatrix} \cos\left((\omega_{NIR} + m\omega_{MIR})t - \frac{(\omega_{NIR} + m\omega_{MIR})\pi}{2(\omega_{NIR} + m\omega_{MIR})}\right) \\ b\sin\left((\omega_{NIR} + m\omega_{MIR})t - \frac{(\omega_{NIR} + m\omega_{MIR})\pi}{2(\omega_{NIR} + m\omega_{MIR})}\right) \end{pmatrix}, \tag{75}$$

where $b$ is a real number. This demonstrates that the sideband light can be elliptical polarization with major or minor axis parallel to the x-axis. In the same way, constraints on the sideband light under x-polarized NIR field can also be determined. Thus, the selection rule for $\widehat{H}_y$ in Table S2 is obtained, which is similar to the selection rules of HHG[4]. Note that the selection rule derived from $\widehat{\sigma}_y$ is stricter than that derived from $\widehat{H}_y$.

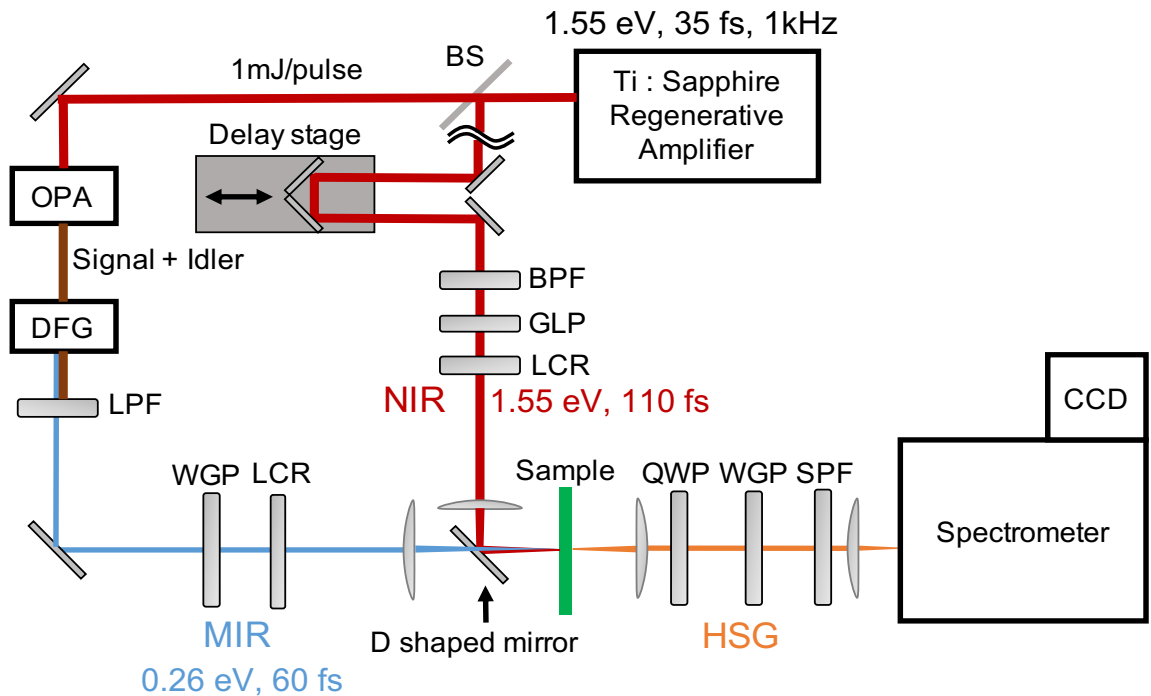

**Figure S4. Experimental Setup for HSG** (BS : beam splitter, BPF : band pass filter : , GLP : Glan laser polarizer, LCR : liquid crystal retarder, OPA : optical parametric amplifier : DFG : difference frequency generation, LPF : long pass filter, WGP：wire grid polarizer, QWP : quarter wave plate, SPF : short pass) filter).

# 6. Estimation of pulse durations

Here, we show how to estimate the pulse duration of the NIR and MIR pulses. We consider gaussian envelops for the MIR and NIR electric field.

$$E_{MIR}(t) = E_M^{(0)} \exp\left(-\frac{t^2}{2\tau_M^2}\right) \tag{76}$$

$$E_{NIR}(t) = E_N^{(0)} \exp\left(-\frac{t^2}{2\tau_N^2}\right) \tag{77}$$

where $E_{MIR}(t)$ and $E_{NIR}(t)$ are the envelops of the MIR and NIR electric field respectively. $\tau_M$ and $\tau_N$ are related to the full width at half maximum of MIR (FWHM$_M$) and NIR (FWHM$_N$) pulses as follows:

$$\text{FWHM}_M = 2\tau_M\sqrt{\ln(2)} \tag{78}$$
$$\text{FWHM}_N = 2\tau_N\sqrt{\ln(2)}. \tag{79}$$

The pulse duration of NIR pulses after bandpass filter and liquid crystal retarder was measured as 110 fs (full width at half maximum) by SPIDER (Spectral Phase Interferometry for Direct Electric-field Reconstruction). We estimate the pulse duration of MIR pulses from time delay dependence of the first-order sideband intensity.

One of the conventional methods to examine an unknown pulse duration is cross-correlation method, where an unknown pulse duration is estimated from time delay dependence of the sum-frequency generation (SFG) in a nonlinear optical media with another known pulse. According to the perturbation theory, the electric field generated by SFG of the NIR and MIR pulses at each time $t'$ is given by

$$E_{SFG}(t, t') \propto E_{MIR}(t - t')E_{NIR}(t'), \tag{80}$$

where the optical response of the media is assumed to be instantaneous. $t$ is the time delay of the NIR and MIR pulses. When the intensity of SFG is measured by a detector which has no time resolution, the integrated intensity $I_{SFG}(t)$ is given by

$$I_{SFG}(t) \propto \int dt' |E_{MIR}(t - t')E_{NIR}(t')|^2 \propto \int dt' I_{MIR}(t - t')I_{NIR}(t'). \tag{81}$$

The MIR pulse duration can be derived from this expression. However, the sideband generation is a nonperturbative phenomenon, Considering excitation power dependence does not follow the power law, we modify the analysis of the cross-correlation method. We approximate the instantaneous response of sideband generation and write the first order sideband intensity as

$$I_S(t) \propto \int dt' f\left(I_{MIR}(t - t')\right) I_{NIR}(t') \tag{82}$$

with a nonlinear function $f$. The function $f$ is determined by fitting the obtained the MIR excitation power dependence of the sideband intensity $F(I_{MIR,0})$, where $I_{MIR,0}$ is the peak power of the MIR pulses written as

$$I_{MIR,0} = \frac{1}{2}c\epsilon_0 E_M^{(0)2}, \tag{83}$$

$c$ is the speed of light and $\epsilon_0$ is the permittivity in a vacuum. We assume that the function $f$ can be approximately written as

$$f(I) = 1 - \exp\left(-\frac{I}{w_1}\right), \tag{84}$$

which is convex and satisfies $f(0) = 0$.

The function $f$ works so that the time duration of $f(I_{MIR}(t))$ becomes longer than $I_{MIR}(t)$. The excitation power dependence $F(I_{MIR,0})$ is measured at time delay 0, therefore that is given by

$$F(I_{MIR,0}) = C \int dt' f(I_{MIR}(-t')) I_{NIR}(t')$$

$$= C' \int dt' \left(1 - \exp\left(-\frac{I_{MIR,0}}{w_1}\exp\left(-\frac{t'^2}{\tau_M^2}\right)\right)\right) \exp\left(-\frac{t'^2}{\tau_N^2}\right), \tag{85}$$

where $C, C'$ is a constant. Furthermore, the time delay dependence $G(t)$ is give by

$$G(t) = C \int dt' f(I_{MIR}(t-t')) I_{NIR}(t')$$

$$= C' \int dt' \left(1 - \exp\left(-\frac{I_{MIR,0,max}}{w_1}\exp\left(-\frac{(t-t')^2}{\tau_M^2}\right)\right)\right) \exp\left(-\frac{t'^2}{\tau_N^2}\right). \tag{86}$$

$I_{MIR,0,max}$ is the maximum MIR power of power dependence, which is used for the measurement of the time delay dependence. The parameters $C', \tau_M, w_1$ are determined through the global fitting of the experimental results.

As a result, $FWHM_M$ is estimated to be 60 fs from $\tau_M$. The fitting curves by eq. (85), (86) are shown as solid lines in Fig. S5.

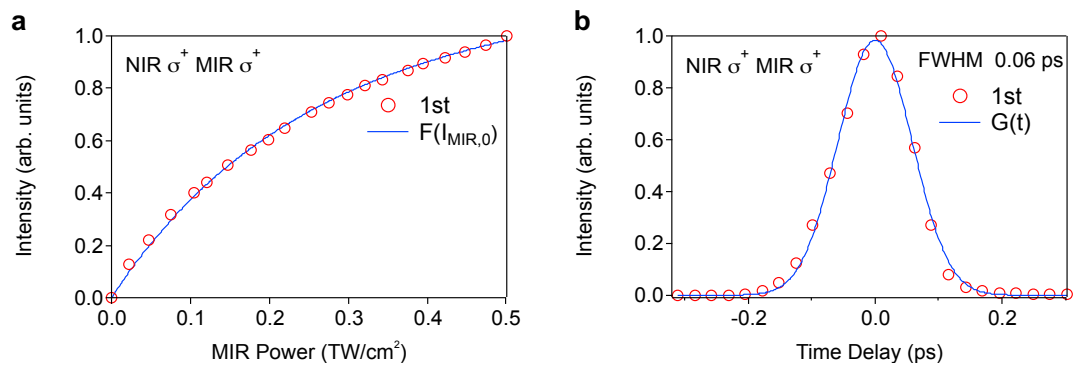

Figure S5. **Estimation of MIR pulse duration. a,b** MIR power dependence (**a**) and time delay dependence (**b**) of the first order sideband intensity under $\sigma^+$-polarized NIR and MIR pulses. The blue solid lines in **a** and **b** are the fitting result with $F(I_{MIR,0})$ and $G(t)$, respectively.

1. Yoshikawa, N., Tamaya, T. & Tanaka, K. High-harmonic generation in graphene enhanced by elliptically polarized light excitation. *Science* **356**, 736–738 (2017).
2. Shirley, J. H. Solution of the schrödinger equation with a hamiltonian periodic in time. *Phys. Rev.* **138**, B979 (1965).
3. Faisal, F. H. M. & Kamiński, J. Z. Floquet-Bloch theory of high-harmonic generation in periodic structures. *Phys. Rev. A* **56**, 748–762 (1997).
4. Neufeld, O., Podolsky, D. & Cohen, O. Floquet group theory and its application to selection rules in harmonic generation. *Nat. Commun.* **10**, 405 (2019).
5. Neufeld, O. *et al.* Ultrasensitive chiral spectroscopy by dynamical symmetry breaking in high harmonic generation. *Phys. Rev. X* **9**, 031002 (2019).
6. Sie, E. J. *et al.* Large, valley-exclusive Bloch-Siegert shift in monolayer $WS_2$. *Science* **355**, 1066–1069 (2017).